\documentclass[12pt,a4paper]{article}
\usepackage{a4wide}
\usepackage{latexsym}
\usepackage{epsf}
\usepackage{amssymb}
\makeatletter
\@addtoreset{equation}{section}
\makeatother

\begin{document}

\begin{flushright}
\small
IFT-UAM/CSIC-99-31\\
{\bf hep-th/0003071}\\
March $9$th, $2000$
\normalsize
\end{flushright}

\begin{center}


\vspace{.7cm}

{\LARGE {\bf Supersymmetry of Topological\\~\\
Kerr-Newman-Taub-NUT-$aDS$ Spacetimes}}

\vspace{1.2cm}

{\bf\large Natxo Alonso-Alberca},${}^{\spadesuit}$
\footnote{E-mail: {\tt Natxo.Alonso@uam.es}}
{\bf\large Patrick Meessen}${}^{\spadesuit}$
\footnote{E-mail: {\tt Patrick.Meessen@uam.es}}
\vskip 0.3truecm
{\bf\large and Tom\'as Ort\'{\i}n}${}^{\spadesuit\clubsuit}$
\footnote{E-mail: {\tt tomas@leonidas.imaff.csic.es}}
\vskip 1truecm

${}^{\spadesuit}$\ {\it Instituto de F\'{\i}sica Te\'orica, C-XVI,
Universidad Aut\'onoma de Madrid \\
E-28049-Madrid, Spain}

\vskip 0.2cm
${}^{\clubsuit}$\ {\it I.M.A.F.F., C.S.I.C., Calle de Serrano 113 bis\\ 
E-28006-Madrid, Spain}

\vspace{.7cm}


{\bf Abstract}

\end{center}

\begin{quotation}

\small

We extend the topological Kerr-Newman-$aDS$ solutions by including
NUT charge and find generalizations of the Robinson-Bertotti solution
to the negative cosmological constant case with different topologies.
We show how all these solutions can be obtained as limits of the 
general Plebanski-Demianski solution.

We study the supersymmetry properties of all these solutions in the
context of gauged $N=2,d=4$ supergravity. Generically they preserve at
most $1/4$ of the total supersymmetry. In the Plebanski-Demianski
case, although gauged $N=2,d=4$ supergravity does not have
electric-magnetic duality, we find that the family of supersymmetric
solutions still enjoys a sort of electric-magnetic duality in which
electric and magnetic charges and mass and Taub-NUT charge are rotated
simultaneously.

\end{quotation}

\newpage

\pagestyle{plain}


\section*{Introduction}

The presence of a negative cosmological constant is enough to
invalidate the classical theorems \cite{art:Haw7,art:FSW} in which it
is proven that at any given time black-hole horizons are always
topologically spheres: asymptotically anti-De Sitter ($aaDS$)
black-hole solutions are known such that the constant-time sections of
their event horizons are not topologically spheres
\cite{art:ABHP,art:Man1,art:Man2,art:SM,art:Le1,art:Le2,art:LeZa,art:CaZh,
  art:HL,art:Bir}.  In particular, $aaDS$ Schwarzschild black holes
with horizons with the topology of Riemann surfaces of arbitrary genus
(henceforth called {\it topological} black holes) were given in
Ref.~\cite{art:Va}, the charged generalization in the framework of the
Einstein-Maxwell theory with a negative cosmological constant
(topological $aaDS$ Reissner-Nordstr\"{o}m (RN-$aDS$) black holes) was
studied in Ref.~\cite{art:BLP}. The generalization to the rotating
case (topological $aaDS$ Kerr-Newman (KN-$aDS$) black holes) was found
and studied in Ref.~\cite{art:KMV} using the general Petrov type D
solution of Plebanski and Demianski (PD solution) Ref.~\cite{art:PD}
(which contains in different limits all these topological black-hole
solutions) and other methods. $aaDS$ black holes with exotic horizons
with topologies are also known in higher dimensions \cite{art:Bir}, in
theories with dilaton \cite{kn:Cai1} and Lovelock gravity
\cite{kn:Cai2}.

The supersymmetry properties of $aaDS$ black holes were first studied
by Romans in the context of $N=2,d=4$ gauged supergravity
\cite{art:Ro} for RN-$aDS$ black holes with spherical horizons.  Later
on, Kosteleck\'y and Perry studied the supersymmetry properties of
KN-$aDS$ black holes \cite{art:KP}.  Recently, Caldarelli and Klemm
extended Romans' results to the case of topological RN-$aDS$ black
holes and extended and corrected Kosteleck\'y and Perry's in the
spherical KN-$aDS$ case in Ref.~\cite{art:CK}.

The supersymmetry properties known are far from being understood.  In
the recent years we have learned how to interpret many supersymmetric
solutions as intersections of ``elementary'' supersymmetric solutions
preserving half of the supersymmetries. Each additional object in the
intersection breaks an additional half of the remaining supersymmetry
\footnote{Except in Hanany-Witten-like cases in which one can add one
  more object to an intersection without breaking any further
  supersymmetry. Needless to say that here we use ``object'' in a
  loose and general way that may include gravitational instantons,
  certain kinds of singularities, etc.}. Thus, in $N=2,d=4$ ungauged
supergravity there is essentially one kind of object which is
point-like and that breaks a half of the available supersymmetry and
one can either break all the supersymmetry or just one half or nothing
at all.

In $N=2,d=4$ gauged supergravity, however, Romans discovered solutions
that preserve just $1/4$ of the supersymmetry, characterized by a
magnetic charge inversely proportional to the coupling constant. The
simplest of those solutions only has magnetic charge (zero mass and
electric charge) equal to the minimal amount of magnetic charge
allowed by Dirac's quantization condition. It is really difficult to
understand this fact using the paradigm of intersection of elementary
objects.

Our goal in this paper is to try to gain some insight into this problem
by examining more general cases an calculating, if possible, the
amount of supersymmetries preserved by the solutions. Thus, in this
letter we first present topological Kerr-Newman-Taub-NUT-$aDS$
solutions and cosmological generalizations of the Robinson-Bertotti
solution and then study their supersymmetry properties together with
those of the general Plebanski-Demianski solution from which all of
them can be obtained through different contractions. We will see that,
generically, these solutions preserve only $1/4$ of the available
supersymmetries in presence of angular momentum. Our second main result
will be the identification of a sort of electric-magnetic duality 
symmetry of the {\it supersymmetric} Plebanski-Demianski solutions
that involves the mass and NUT charge.

This paper is organized as follows: in Section~\ref{sec-N2} we
describe $N=2,d=4$ gauged Supergravity. In Section~\ref{sec-solutions}
we describe the solutions whose supersymmetry properties we are going
to study. In Section~\ref{sec-susy-KN-TN} we study the integrability
conditions of the Killing spinor equation for the topological
KN-TN-$aDS$ solutions. In Section~\ref{sec-susy-RB} and
Section~\ref{sec-susy-PD} we perform the same analysis for RB-$aDS$
and the general PD solutions respectively.
Section~\ref{sec-conclusions} contains our conclusions.


\section{Cosmological EM Theory and $N=2,d=4$ Gauged
Supergravity}
\label{sec-N2}

The $N=2,d=4$ supergravity multiplet consists of the Vierbein, a couple
of real gravitini and a vector field

\begin{equation}
\{e_{\mu}{}^{a},\psi_{\mu}=
\left(
\begin{array}{c}
\psi^{1}_{\mu}\\
\psi^{2}_{\mu}\\
\end{array}
\right),A_{\mu}\}\, ,
\end{equation}

\noindent respectively. With this multiplet one can construct two
different supergravity theories: ``pure'' $N=2,d=4$ supergravity and
``gauged'' $N=2,d=4$ supergravity. The former can be understood as the
zero-coupling limit of the latter and the second as the theory one
obtains by gauging the $SO(2)$ symmetry that rotates the gravitini.
The gauged $N=2,d=4$ supergravity action for these fields in the 1.5
formalism is

\begin{equation}
\begin{array}{rcl}
S_{g} & = & {\displaystyle\int} d^{4}x\, e \left\{ R(e,\omega) +6g^{2}
+2 e^{-1}\epsilon^{\mu\nu\rho\sigma}\bar{\psi}_{\mu}\gamma_{5}\gamma_{\nu}
\left(\hat{\cal D}_{\rho}+ig A_{\rho}\sigma^{2}\right)\psi_{\sigma}
-{\cal F}^{2} \right.\\
& & \\
& & 
\left. 
+{\cal J}_{(m)}{}^{\mu\nu}
({\cal J}_{(e) \mu\nu} +{\cal J}_{(m) \mu\nu}) \right\}\, ,
\end{array}
\end{equation}

\noindent where  $\hat{\cal D}$ is the $SO(2,3)$ gauge 
covariant derivative

\begin{equation}
\hat{\cal D}_{\mu} = {\cal D}_{\mu} 
-{\textstyle\frac{i}{2}}g\gamma_{\mu}\, ,
\end{equation}

\noindent  $F$ is the standard vector field strength, 
$\tilde{F}$ is the supercovariant field strength and we also 
define for convenience ${\cal F}$ by

\begin{equation}
\left\{
\begin{array}{rcl}
F_{\mu\nu} & = & 2\partial_{[\mu} A_{\nu]}\, ,\\
& & \\
\tilde{F}_{\mu\nu} & = & F_{\mu\nu} +{\cal J}_{(e) \mu\nu}\, ,\\
& & \\
{\cal F}_{\mu\nu} & = &  \tilde{F}_{\mu\nu} +{\cal J}_{(m) \mu\nu}\, , \\
\end{array}
\right.
\end{equation}

\noindent where we have also defined

\begin{equation}
\left\{
\begin{array}{rcl}
{\cal J}_{(e) \mu\nu} & = & i\bar{\psi}_{\mu}\sigma^{2}\psi_{\nu}\, ,\\
& & \\
{\cal J}_{(m) \mu\nu} & = & -\frac{1}{2e}\epsilon^{\mu\nu\rho\sigma}
\bar{\psi}_{\rho}\gamma_{5}\sigma^{2}\psi_{\sigma}\, .\\
\end{array}
\right.
\end{equation}

We see that the gauge coupling constant $g$ is related to the
cosmological constant by

\begin{equation}
\Lambda=-3g^{2}\, .  
\end{equation}

The equation of motion for $\omega_{\mu}{}^{ab}$ implies that it is
given by 

\begin{equation}
\label{eq:torsionfulspinconnection}
\left\{
\begin{array}{rcl}
\omega_{abc} & = & 
-\Omega_{abc} + \Omega_{bca} - \Omega_{cab}\, ,\\
& & \\
\Omega_{\mu\nu}^{a} & = & \Omega_{\mu\nu}{}^{a} (e) 
+\frac{1}{2}T_{\mu\nu}{}^{a}\, ,\\
& & \\
\Omega_{abc}(e) & = & 
e^{\mu}{}_{a}e^{\nu}{}_{b} \partial_{[\mu}e_{\nu]c}\, ,\\
& & \\
T_{\mu\nu}{}^{a} & = & i\bar{\psi}_{\mu}\gamma^{a}\psi_{\nu}\, .\\
\end{array}
\right.
\end{equation}

\noindent It is assumed that this equation has been used everywhere 
(1.5 formalism).

The Maxwell equation and Bianchi identity are

\begin{equation}
\left\{
\begin{array}{rcl}
\partial_{\mu}(e\, {\cal F}^{\mu\nu}) 
& = & {\textstyle\frac{ig}{2}}\epsilon^{\nu\lambda\rho\sigma}
\bar{\psi}_{\lambda}\gamma_{5}\gamma_{\rho}\sigma^{2}\psi_{\sigma}\, ,\\
& & \\
\partial_{\mu}(e\, {}^{\star} F^{\mu\nu}) & = & 0\, .\\
\end{array}
\right.
\end{equation}

\noindent Observe that the divergences of ${\cal J}_{e}$ and 
${\cal J}_{m}$ are two topologically conserved currents that appear as
electric-like and magnetic-like sources for the vector field in the
Maxwell equation

\begin{equation}
\partial_{\mu}(eF^{\mu\nu})= 
+\partial_{\mu}(e {\cal J}_{e}^{\nu\mu})
+\partial_{\mu}(e{\cal J}_{m}^{\nu\mu})
+ {\textstyle\frac{ig}{2}}\epsilon^{\nu\lambda\rho\sigma}
\bar{\psi}_{\lambda}\gamma_{5}\gamma_{\rho}\sigma^{2}\psi_{\sigma}\, .  
\end{equation}

\noindent  They are naturally associated to the electric and 
magnetic central charges of the $N=2,d=4$ supersymmetry algebra.  The
third term in the r.h.s.~of the above equation is associated to the
gravitino electric charge and it is, therefore, proportional to the
gauge constant. Finally, the Einstein and Rarita-Schwinger equations
are

\begin{equation}
\left\{
\begin{array}{rcl}
0 & = & G_{a}{}^{\mu} -3g^{2} e_{a}{}^{\mu} -2T(\psi)_{a}{}^{\mu}
-2\tilde{T}(A)_{a}{}^{\mu}\, ,\\
& & \\
0 & = & e^{-1}\epsilon^{\mu\nu\rho\sigma}
\gamma_{5}\gamma_{\nu}
\left(\hat{\cal D}_{\rho}+ig A_{\rho}\sigma^{2}\right)\psi_{\sigma}
-i\left(\tilde{F}^{\mu\nu} +i{}^{\star}\tilde{F}^{\mu\nu}\gamma_{5}
\right)\sigma^{2}\psi_{\nu}\, ,\\
\end{array}
\right.
\end{equation}

\noindent where the equation of motion for $\omega_{\mu}{}^{ab}$ has
been used and where

\begin{equation}
\label{eq:energymomentumtensors}
\left\{
\begin{array}{rcl}
T(\psi)_{a}{}^{\mu} & = & -\frac{1}{2e}\epsilon^{\mu\nu\rho\sigma}
\bar{\psi}_{\nu}\gamma_{5}\gamma_{a}
\left(\hat{\cal D}_{\rho}+ig A_{\rho}\sigma^{2}\right)\psi_{\sigma}\\
& & \\
& & -\frac{ig}{4e} \epsilon^{\mu\nu\rho\sigma}
\bar{\psi}_{\nu}\gamma_{5}\gamma_{\rho a}\psi_{\sigma}\, ,\\
& & \\
\tilde{T}(A)_{a}{}^{\mu} & = & 
\tilde{F}_{a}{}^{\rho}\tilde{F}^{\mu}{}_{\rho}
-\frac{1}{4}e_{a}{}^{\mu}\tilde{F}^{2}\, .\\
\end{array}
\right.
\end{equation}

Apart from invariance under general coordinate and local Lorentz
transformations the theory is invariant under $U(1)$ gauge
transformations

\begin{equation}
\left\{
\begin{array}{rcl}
A^{\prime}_{\mu} & = & A_{\mu}+\partial_{\mu}\chi\, ,\\
& & \\
\psi_{\mu}^{\prime} & = & e^{-ig\chi\sigma^{2}} \psi_{\mu}\, ,\\
\end{array}
\right.
\end{equation}

\noindent and local $N=2$ supersymmetry transformations

\begin{equation}
\label{eq:susyt}
\left\{
\begin{array}{rcl}
\delta_{\epsilon} e_{\mu}{}^{a} & = & 
-i\bar{\epsilon}\gamma^{a}\psi_{\mu}\, ,\\
& & \\
\delta_{\epsilon} A_{\mu} & = & 
-i\bar{\epsilon}\sigma^{2}\psi_{\mu}\, ,\\
& & \\
\delta_{\epsilon} \psi_{\mu} & = & 
\tilde{\hat{\cal D}}_{\mu}\epsilon\, ,\\
\end{array}
\right.
\end{equation}

\noindent where the $\tilde{\hat{\cal D}}_{\mu}$ is the
 supercovariant derivative defined by

\begin{equation}
\label{eq:supercovariantderivative}
\tilde{\hat{\cal D}}_{\mu}=
\hat{\cal D}_{\mu} +igA_{\mu}\sigma^{2}
+{\textstyle\frac{1}{4}}\not\!\tilde{F}\gamma_{\mu}\sigma^{2}\, .
\end{equation}

In the ungauged case, the theory enjoys {\it chiral-dual} invariance
which interchanges the Maxwell and Bianchi equations and the
topologically conserved electric and magnetic charges (and, therefore,
the associated central charges). In the gauged theory, the gauge
coupling breaks this invariance.

We are going to work with purely bosonic solutions of this theory.
They obey the bosonic equations of motion

\begin{equation}
\left\{ 
\begin{array}{rcl}
\nabla_{\mu}F^{\mu\nu} & = & 0\, ,\\
& & \\
\nabla_{\mu}{}^{\star}F^{\mu\nu} & = & 0\, ,\\
& & \\
R_{\mu\nu} & = & 2T_{\mu\nu}(A) - 3g^{2} g_{\mu\nu} \, ,\\
\end{array}
\right.
\end{equation}

\noindent where $T_{\mu\nu}(A)$ is just the standard 
energy-momentum tensor for an Abelian gauge field:

\begin{equation}
T_{\mu\nu}(A) = F_{\mu}{}^{\rho}F_{\rho\nu} 
-{\textstyle\frac{1}{4}} g_{\mu\nu}F^{2}\, .  
\end{equation}

These equations of motion are duality-invariant. However, the
gravitino supersymmetry rule (even with fermionic fields set to zero)
is not duality-invariant and the supersymmetry properties of
duality-related bosonic solutions are not, in general, the same.


\section{Topological RN-TN-$aDS$, KN-TN-$aDS$ and 
RB-$aDS$ and PD Solutions}
\label{sec-solutions}

In this section we display and describe the solutions whose
supersymmetry properties will later be studied.  For simplicity we
start with the unrotating RN-TN-$aDS$ although they are included in
the general KN-TN-$aDS$ case.


\subsection{Topological RN-TN-$aDS$ Solutions}

These solutions generalize, by including NUT charge $N$, the
topological RN-$aDS$ black hole solutions found in Ref.~\cite{art:BLP}.
There are three cases labeled by the parameter $\aleph$ whose value is
essentially the sign of one minus the genus of the horizon and
therefore takes the values $1,0,-1$ for the sphere (genus zero), the
torus (genus 1) and higher genus Riemann surfaces, respectively.  In
the three cases the metric can be written in this form

\begin{equation}
\label{eq:RNTNaDS}
\left\{
\begin{array}{rcl}
ds^{2} & = & {\displaystyle\frac{\lambda}{R^{2}}}
\left( dt +\omega_{\aleph}d\varphi\right)^{2}
-{\displaystyle\frac{R^{2}}{\lambda}} dr^{2} 
-R^{2}d\Omega_{\aleph}^{2}\, ,\\
& & \\
\lambda & = & {\displaystyle\left[g^{2}R^{4} 
+(\aleph +4g^{2}N^{2})(r^{2}-N^{2})-2Mr +|Z|^{2}\right]}\, ,\\
& & \\
R^{2} & = & r^{2}+N^{2}\, ,\\
\end{array}
\right.
\end{equation}

\noindent where $d\Omega_{\aleph}^{2}$ is the metric of the unit sphere, 
the plane and the upper half plane respectively

\begin{equation}
d\Omega^{2}_{\aleph} =
\left\{
\begin{array}{lrcl}
d\theta^{2} +\sin^{2}{\theta}d\varphi^{2}\, ,\hspace{1cm}&
\aleph & = & +1\, ,\\
& & & \\
d\theta^{2} +d\varphi^{2}\, ,& \aleph & = & 0\, ,\\
& & & \\
d\theta^{2} +\sinh^{2}{\theta}d\varphi^{2}\, ,&
\aleph & = & -1\, ,\\
\end{array}
\right.
\end{equation}

\noindent $\omega_{\aleph}$ is the function

\begin{equation}
\omega_{\aleph} =
\left\{
\begin{array}{lrcl}
2N\cos{\theta}\, ,\hspace{1.2cm} & \aleph & = & +1\, ,\\
& & & \\
-2N\theta\, ,& \aleph & = & 0\, ,\\
& & & \\
-2N\cosh{\theta}\, , & \aleph & = & -1\, ,\\
\end{array}
\right.
\end{equation}

\noindent and the vector potential is given by

\begin{equation}
  \begin{array}{rcl}
A_{t} & = & \left( Qr-NP\right) /R^{2}\, ,\\
& & \\
A_{\varphi} & = & 
\left\{
\begin{array}{lcrcl}
\cos{\theta}{\displaystyle\left[P(r^{2}-N^{2}) +2NQr\right]}/R^{2}\, ,
\hspace{.5cm} & {\rm for} & \aleph & = & +1\, ,\\  
 & & & & \\
-\theta {\displaystyle\left[P(r^{2}-N^{2}) +2NQr\right]}/R^{2}\, ,
\hspace{.5cm} & {\rm for} & \aleph & = & 0\, ,\\  
 & & & & \\
-\cosh{\theta}{\displaystyle\left[P(r^{2}-N^{2}) +2NQr\right]}/R^{2}\, ,
\hspace{.5cm} & {\rm for} & \aleph & = & -1\, .\\  
\end{array}
\right.
\end{array}
\end{equation}

It is understood that one has to take the equation of the last two
spacetimes by a discrete group in order to get a torus or a Riemann
surface of arbitrary genus.

These solutions are valid in the $g=0$ case. In that limit (with $N=0$)
so we can speak of black holes, only the $\aleph=+1$ ones can have a
regular event horizon, in agreement with \cite{art:Haw7,art:FSW}.
With $g\neq 0$ (still with $N=0$) and we recover the solutions of
Ref.~\cite{art:BLP} in which $M$ is the mass, $Q$ the electric charge,
$P$ the magnetic charge and $Z=Q+iP$ some of which are black holes
with regular horizons of different topologies.

For $g=0,N\neq 0$ we recover the standard RN-TN solutions in which
those parameters are still the physical parameters\footnote{A
  definition of the mass of Taub-NUT spaces cannot be given in the
  standard form because these solutions do not go asymptotically to
  any other vacuum solution.  The same happens in the 5-dimensional KK
  monopole solution, studied in Refs.~\cite{kn:DS,kn:BKKLS}. However,
  as different from the KK monopole, the TN solution is not
  ultrastatic and the tricks used in those references to define and
  calculate the mass of the KK monopole do not seem to apply to this
  case. A definition inspired in the AdS/CFT correspondence, has,
  however, been recently given in Refs.~\cite{kn:Man3,kn:CEJM}.} and
$N$ is the NUT charge. When the product $gN\neq 0$ it is no longer
clear that $M,Q,P$ are the true mass, electric and magnetic charges
that appear in the superalgebra.  This is similar to what happens in
the rotating case \cite{art:KP} in which the true charges are
combinations of the parameters $M,P,Q$ appearing in the solution with
the product $ga$.

It is useful to have a general form of the solutions valid for the
three cases $\aleph=1,0,-1$. To have such a general expression we
define the coordinate $u$

\begin{equation}
 u \;\equiv\; 
\left\{
\begin{array}{lrcl}
-\cos{\theta}\, ,\hspace{1cm} & \aleph & = &  +1\, , \\
& & & \\
\theta\, , & \aleph & = & 0\, , \\
& & & \\
\cosh{\theta}\, ,  & \aleph & = & -1\, , 
\end{array}
\right.
\end{equation}

\noindent and then 

\begin{equation}
\left\{
\begin{array}{rcl}
ds^{2} & = & 
{\displaystyle\frac{\lambda}{R^{2}}}
\left(dt -2Nud\varphi\right)^{2} 
-{\displaystyle\frac{R^{2}}{\lambda}}dr^{2}
-{\displaystyle\frac{R^{2}}{S(u)}}du^{2} -R^{2}\, S(u)d\varphi^{2} \; ,\\
& & \\
A_{t} & = & \left( Qr-NP\right) /R^{2}\, ,\\
& & \\
A_{\varphi} & = & 
-u \left[ P\left( r^{2}-N^{2}\right)+ 2NrQ \right]/R^{2} \; ,\\
& & \\
S(u) & = &  \aleph (1-u^{2})+1-\aleph^{2}\, ,\\
\end{array}
\right.
\end{equation}

\noindent where $\lambda$ and $R$ are as above.


\subsection{Topological KN-TN-$aDS$ Solutions}

These solutions generalize the topological KN-$aDS$ solutions given in
Ref.~\cite{art:KMV,art:CK} to the non-zero NUT charge case.  In the
$t,r,u,\varphi$ coordinate system (which is Boyer-Lindquist-type) they
can be written as follows:

\begin{equation}
\label{eq:KNTNaDS}
\left\{
\begin{array}{rcl}
ds^{2} & = & 
{\displaystyle\frac{\lambda}{R^{2}(r,u)}
\left\{dt-\left[2Nu
-a\left(\aleph^{2}-u^{2}\right)\right]d\varphi\right\}^{2}}
-{\displaystyle\frac{R^{2}(r,u)}{\lambda}} dr^{2} \\
& & \\
& &   
-{\displaystyle\frac{R^{2}(r,u)}{\mathcal{S}(u)}}du^{2}
-{\displaystyle\frac{\mathcal{S}(u)}{R^{2}(r,u)}
\left[\left(r^{2}+N^{2}+\aleph^{2}a^{2}\right)d\varphi 
                +adt \right]^{2}} \; . \\
& & \\
A_{t} & = & {\displaystyle\left[Qr-P(N+au)\right]}/R^{2}(r,u)\, ,\\
& & \\
A_{\varphi} & = & {\displaystyle\frac{1}{a}}
\sqrt{r^{2}+N^{2}+\aleph^{2}a^{2}}\,\,
{\displaystyle\left[Qr-P(N+au)\right]}/R^{2}(r,u)\, ,\\
& & \\
\lambda & = &  g^{2}r^{4} 
+\left(\aleph +\aleph^{2}a^{2}g^{2} +6g^{2}N^{2}\right) r^{2}
-2Mr +|Z|^{2} \\
& & \\
& &  
-N^{2}\left(\aleph -3\aleph^{2}a^{2}g^{2} +3g^{2}N^{2}\right)
+a^{2}\left(1+\aleph -\aleph^{2}\right)\; ,\\
& & \\
\mathcal{S}(u) & = & S(u)
+\left(a^{2}g^{2}\, u^{2}\,+\, 4ag^{2}N\, u\right)
\left( u^{2}-\aleph^{2} \right)\; , \\
& & \\
R^{2}(r,u) & = & r^{2}\,+\, \left(N\,+\, au\right)^{2}\; ,\\
\end{array}
\right.
\end{equation}

\noindent with $S(u)$ as above.

In Appendix~\ref{sec-KNTNADSfromPD} it is explained how this metric
can be obtained from the general solution of Plebanski and Demianski
\cite{art:PD}. The above form of the potentials is valid for $a\neq
0$. The $a\rightarrow 0$ limit of the field strength is perfectly well
defined.


\subsection{Topological RB Solutions}

In ungauged $N=2,d=4$ Supergravity, the extremal RN black hole can be
seen as a soliton interpolating between two supersymmetric vacua:
Minkowski spacetime at infinity and RB in the near-horizon limit. The
RB spacetime is the product $aDS_{2}\times S^{2}$ where both factors
are maximally symmetric spaces with opposite curvatures that cancel
each other.  The same thing occurs with other $p$-branes in higher
dimensions \cite{art:GibTown,kn:BST} where the role of the RB spacetime
is played  by $aDS_{p+2}\times S^{8-p}$.  Here we present a
generalization of the RB spacetime to the case of gauged $N=2,d=4$
Supergravity (cosmological E-M theory) whose supersymmetry properties
we will study later. They are the product of $aDS_{2}$ with a sphere
$S^{2}$, a torus $T^{2}$ or a higher-genus Riemann surface
$\Sigma_{g}$ in which now the curvature of the $aDS_{2}$ spacetime is
not completely canceled by the other factor space but they add up to
the 4-dimensional cosmological constant

\begin{equation}
\left\{
\begin{array}{rcl}
ds^{2} & = & K^{2}r^{2}dt^{2}
\,-\,{\displaystyle\frac{1}{K^{2}r^{2}}}dr^{2}
\,-\, L^{-2}S(u)^{-1}du^{2} \,-\, L^{-2}S(u)d\varphi^{2} \; , \\
& & \\
F_{01} & = &  \alpha\, \\
& & \\
F_{23} & = &  -\beta \, ,\\
\end{array}
\right.
\end{equation}

\noindent  where the constants $K,L,\alpha,\beta$ satisfy

\begin{equation}
\begin{array}{rcl}
g^{2} & = & 
\textstyle{\frac{1}{6}}\left\{K^{2}\;-\; \aleph L^{2} \right\}\; , \\
& & \\
\alpha^{2}\,+\, \beta^{2} & = & 
\textstyle{\frac{1}{2}}\left( K^{2}\; +\; \aleph L^{2}\right) \; .\\
\end{array}
\label{eq:RobBertEis}
\end{equation}

\noindent The field strength is covariantly constant  and in this 
coordinate system has constant components which correspond to the
vector potential components

\begin{equation}
\left\{
\begin{array}{rcl}
A_{t} & = & -\alpha r\, ,\\
& & \\
A_{\varphi} & = & -\beta/L^{2} u\, .\\
\end{array}
\right.
\end{equation}

The $\aleph=-1, K^{2}=2L^{2}$ solution, which has special
supersymmetry properties has been also given in \cite{kn:CKZ}.


\subsection{PD Solutions}

Plebanski and Demianski found {\it most general Petrov type D}
solution of the cosmological E-M theory. This general solution
contains as limiting cases all the known solutions, and, in particular
the topological KN-TN-$aDS$ solutions presented above (which in their
turn, also contain the RN-TN-$aDS$ solutions presented at the
beginning). This is shown in Appendix~\ref{sec-KNTNADSfromPD}.

The PD solution depends on the constants\footnote{These constants are
  different from the constants $M,N,Q,P$ that appear in the previous
  solutions.}
$\mathsf{M},\mathsf{N},\mathsf{Q},\mathsf{P},\mathsf{E},\alpha$ and,
of course, $g$, and, in Boyer coordinates $\tau,\sigma,p,q$, reads
\cite{art:PD}

%
%
%
%

\begin{equation}
\label{eq:D&P2metric}
\left\{
\begin{array}{rcl}
ds^{2} & = &  {\displaystyle\frac{\mathcal{Q}(q)}{p^{2}+q^{2}}
              \left( d\tau -p^{2}d\sigma \right)^{2}
            -\frac{p^{2}+q^{2}}{\mathcal{Q}(q)} dq^{2}
            -\frac{p^{2}+q^{2}}{\mathcal{P}(p)} dp^{2}
            -\frac{\mathcal{P}(p)}{p^{2}+q^{2}}
              \left( d\tau +q^{2}d\sigma \right)^{2}} \; ,\\
& & \\
F_{01} & = & (q^{2}+p^{2})^{-2}
\left[\mathsf{Q}(q^{2}-p^{2})-2\mathsf{P}pq\right] \, ,  \\  
& & \\
F_{23} &=& -(q^{2}+p^{2})^{-2}
\left[\mathsf{P}(q^{2}-p^{2})+2\mathsf{Q}pq\right] \, , \\
& & \\
\mathcal{Q}(q) & = & g^{2}q^{4}+\mathsf{E}q^{2}-2\mathsf{M}q
                   +\mathsf{Q}^{2}+\alpha \, , \\
& & \\
\mathcal{P}(p) & = & g^{2}p^{4}-\mathsf{E}p^{2}+2\mathsf{N}p
                   -\mathsf{P}^{2}+\alpha \, .\\
\end{array}
\right.
\end{equation}

This class of solutions has a scaling invariance given by

\begin{equation}
\begin{array}{rclrclrcl}
q          & \rightarrow & \kappa q\, ,              &
\mathsf{M} & \rightarrow & \kappa^{3}\mathsf{M}\, ,\hspace{1cm}  & 
\mathsf{N} & \rightarrow & \kappa^{3}\mathsf{N}\, ,  \\
& & & & & & & & \\
p          & \rightarrow & \kappa p\, ,              & 
\mathsf{Q} & \rightarrow & \kappa^{2}\mathsf{Q}\, ,  & 
\mathsf{P} & \rightarrow & \kappa^{2}\mathsf{P}\, ,\hspace{1cm}  \\
& & & & & & & & \\
\tau       & \rightarrow & \kappa^{-1}\tau\, ,\hspace{1cm}       & 
\mathsf{E} & \rightarrow & \kappa^{2}\mathsf{E}\, ,  &
\alpha     & \rightarrow & \kappa^{4}\alpha\, ,      \\
& & & & & & & & \\
\sigma     & \rightarrow & \kappa^{-3} \sigma\, ,    & 
& & & & & \\
\end{array}
\label{eq:PDscaling}
\end{equation}

\noindent which can be used to bring one of the charges to a given value. 
It is clear that this scaling freedom remains if one of the charges
happens to be nil.

The curvature is determined by $\mathsf{M}$, $\mathsf{N}$,
$\mathsf{Q}$ and $\mathsf{P}$ and one can see that when they are zero,
the Weyl tensor vanishes.  This then means that in that case, the
solution is locally $aDS_{4}$.


\section{Supersymmetry and Integrability Conditions}
\label{sec-susy-KN-TN}

The bosonic part of the supercovariant derivative for gauged $N=2$
supergravity is

\begin{equation}
\tilde{\hat{\cal D}}_{\mu}=
\hat{\nabla}_{\mu}  +gA_{\mu}i\sigma^{2}
-{\textstyle\frac{i}{4}}\not\!F\gamma_{\mu}i\sigma^{2}\, ,
\end{equation}

\noindent where $\hat{\nabla}_{\mu}$ is the $SO(2,3)$ 
gauge-covariant derivative.  The Killing spinor equation is

\begin{equation}
\tilde{\hat{D}}_{\mu}\epsilon=0\, ,
\end{equation}

\noindent and a necessary condition for it to have solutions is the
integrability condition

\begin{equation}
\left[\tilde{\hat{\cal D}}_{\mu},\tilde{\hat{\cal D}}_{\nu}\right]  
\epsilon=0\, .
\end{equation}

\noindent One finds \cite{art:Ro}

\begin{equation}
\label{eq:integrability}
\begin{array}{rcl}
\left[\tilde{\hat{\cal D}}_{\mu},\tilde{\hat{\cal D}}_{\nu}\right]  
\epsilon
& = & -{\textstyle\frac{1}{4}} 
\left\{ C_{\mu\nu}{}^{ab} \gamma_{ab}
+2i\not\!\nabla \left(F_{\mu\nu} +i\, {}^{\star}F_{\mu\nu}\gamma_{5} \right)
i\sigma^{2}\right.\\
& & \\
& & 
\left.
+{\textstyle\frac{g}{2}} F_{ab}\left(3\gamma^{ab}\gamma_{\mu\nu}
+\gamma_{\mu\nu}\gamma^{ab}\right)i\sigma^{2}
\right\}\epsilon=0\, .\\
\end{array}
\end{equation}

We study first the non-rotating case RN-TN-$aDS$ case.


\subsection{Supersymmetry of Topological RN-TN-$aDS$ Solutions}

Introducing the Vierbein 1-forms

\begin{equation}
\begin{array}{rclrcl}
e^{0} & = & \lambda^{1/2}/R\left( dt +\omega_{\aleph}d\varphi \right)\, ,
\hspace{.5cm} & 
e^{1} & = & \lambda^{-1/2} Rdr \, , \\
& & & & & \\
e^{2} & = & R\, d\theta\, , & 
e^{3} & = &  R\, \Omega_{\aleph}\, d\varphi \, ,\\
\end{array}
\end{equation}

\noindent we find

\begin{equation}
\begin{array}{rcl}
F_{01} & = & \left(Q(r^2 -N^2 )-2NP r\right)/R^{4} \, , \\
& &  \\
F_{23} & = & -\left(P(r^2 -N^2 ) + 2NQ r\right)/R^{4} \, ,\\
\end{array}
\end{equation}

\begin{equation}
\begin{array}{rcl}
\nabla_{1}F_{01} & = & -2\lambda^{1/2}/ R^{7} \left[ Q(r^{3}-3rN^{2})
-P(3r^{2}N -N^{3})\right]\, ,\\
& & \\
\nabla_{1}{}^{\star}F_{01} & = & -2\lambda^{1/2}/ R^{7} \left[ P(r^{3}-3rN^{2})
+Q(3r^{2}N -N^{3})\right]\, ,\\
\end{array}
\end{equation}

\noindent (the remaining components of $\nabla_{a}F_{bc}$ can be
found using the Bianchi identities or the Maxwell equations, which are
satisfied) and

\begin{equation}
\begin{array}{rcl}
-\frac{1}{2}C_{01}{}^{01} & = & C_{02}{}^{02} = C_{03}{}^{03} =
C_{12}{}^{12} =  C_{13}{}^{13} = -\frac{1}{2}C_{23}{}^{23} =   C_{1}\, ,\\
& & \\
C_{02}{}^{13} & = & -C_{03}{}^{12} = C_{12}{}^{03} = -C_{13}{}^{02}  = 
-\frac{1}{2}C_{23}{}^{01} =  C_{2}\, ,\\
& & \\
C_{1} & = & \left[Mr^{3} -\left(3N^{2}(\aleph-4g^{2}N^{2}) 
+|Z|^{2}\right)r^{2}\right.\\
& & \\
& & 
\left.
-3N^{2}Mr +N^{2}\left(N^{2}(\aleph-4g^{2}N^{2}) 
+|Z|^{2}\right)\right]/R^{6}\, ,\\
& & \\
C_{2} & = & -N\left[(\aleph-4g^{2}N^{2}) r^{3} +3Mr^{2}\right.\\
& & \\
& & 
\left.
-\left(3N^{2}(\aleph-4g^{2}N^{2}) +2|Z|^{2}\right)r -MN^{2}\right]/R^{6}\, ,\\
\end{array}
\end{equation}

Plugging all this into the integrability conditions we get the following 
conditions on the parameters:

\begin{eqnarray}
0 &=&  g\left[ MP \,+\, QN(\aleph +4g^2 N^2) \right] \, , \\
& & \nonumber \\
0 &=& {\cal B}_{+}{\cal B}_{-}\, ,\\
\end{eqnarray}

\noindent where we have defined

\begin{equation}
{\cal B}_{\pm} \equiv(M\mp gNQ)^2 +N^2 (\aleph \pm gP +4g^2 N^2 )^{2}
      -(\aleph \pm 2gP + 5 g^2 N^2 )|Z|^2\, .
\end{equation}

The first condition plays the role of a constraint which is
automatically satisfied in the well-known $g=0$ case, while the second
implies ${\cal B}_{\pm}=0$ which should be the (saturated) Bogomol'nyi
bound of gauged $N=2,d=4$ supergravity and actually it reduces to the
well-known Bogomol'nyi bound of ungauged $N=2,d=4$ supergravity in
asymptotically flat spaces ($\aleph=+1$) generalized so as to include
NUT charge (see Refs.~\cite{art:KKOT,art:BKO3,art:AMO})
$M^{2}+N^{2}=Q^{2}+P^{2}$. For $g=0$ and arbitrary $\aleph$ we get

\begin{equation}
M^{2} +\aleph^{2}N^{2}= \aleph (Q^{2}+P^{2})\, .  
\end{equation}

A detailed analysis of the different cases in which the constrains is
satisfied and the Bogomol'nyi bound is saturated gives as a result the
four cases represented in Table~\ref{tab-RNTNADS}. 

The first corresponds evidently to $aDS_{4}$ itself in standard
spherical coordinates, which is maximally supersymmetric and
preserves all supersymmetries. The second case can be shown to
describe, at least locally, $aDS_{4}$ as well (the Weyl tensor
vanishes and the space is maximally symmetric). There are, thus, two
different values of the parameter $N$ that correspond to the same
spacetime.

In the third and fourth cases we have taken for the sake of
convenience $Q$ and $N$ as independent parameters.  The third case is
a generalization to $gN\neq 0$ of the standard $M=|Q|$ case of
ungauged $N=2,d=4$ supergravity where $Q$ is arbitrary which preserves
$1/2$ of the supersymmetries. Here a non-vanishing magnetic charge
proportional to $N$ is induced.  As a matter of fact, it admits the
limits $g\rightarrow 0$ and/or $N\rightarrow 0$ with the same amount
of supersymmetry preserved. 

There are two particularly interesting limits: the often neglected
$g=0,\aleph=0$ case which (setting $N=0$ for simplicity and rescaling
the coordinates $\theta,\varphi$ which do not represent angles
anymore) corresponds to the solution

\begin{equation}
\left\{
\begin{array}{rcl}
ds^{2} & = & {\displaystyle\frac{Q^{2}}{r^{2}}} dt^{2}
- {\displaystyle\frac{r^{2}}{Q^{2}}}
\left(dr^{2} +d\theta^{2} +d\varphi^{2} \right)\, ,\\
& & \\
A_{t} & = & {\displaystyle \frac{Q}{r}}\, .\\
\end{array}
\right.
\end{equation}

This solution belongs to the Papapetrou-Majumdar class

\begin{equation}
\left\{
\begin{array}{rcl}
ds^{2} & = & H^{-2}dt^{2} -H^{2}d\vec{x}^{\, 2}\, ,\\
& & \\
A_{t} & = & \pm H^{-1}\, ,\\
& & \\
\partial_{\underline{i}}\partial_{\underline{i}} H & = & 0\, ,\\
\end{array}
\right.
\end{equation}

\noindent where the harmonic function $H$ has been chosen to depend on
only one coordinate $H=|Q|x$ and not on $y,z$.

The second interesting limit $Q\rightarrow 0$ also gives a
supersymmetric configuration that preserves $1/2$ of the
supersymmetries with only magnetic and NUT charge and zero mass.

The fourth case in Table~\ref{tab-RNTNADS} preserves only $1/4$ of the
supersymmetries and only exists for $g\neq 0$. It is a generalization
to $N\neq 0$ of Romans' global monopole solution \cite{art:Ro}. We see that the
presence of both NUT and electric charge implies that the mass
parameter has to be finite. On the other hand, it admits the limits
$Q\rightarrow 0$ and/or $N\rightarrow 0$ with the same amount of
supersymmetry preserved.

\begin{table}
\footnotesize
\begin{center}
\begin{tabular}{||c|c|c|c|c|c||}
\hline\hline
& & & & & \\
$M$ & $N$ &  $Q$ & $P$ & $\aleph$ & SUSY \\ 
\hline\hline
& & & & & \\
0 & 0  &    0                &  0    & $+1$ & 1     \\ 
& & & & & \\
\hline
& & & & & \\
0    & $\pm {\displaystyle\frac{1}{2g}}$ &  0     &  0   &  $-1$ &   1   \\ 
& & & & & \\
\hline
& & & & & \\ 
$|Q\sqrt{\aleph +4g^{2}N^{2}}|$ & any &  any  & $\pm N\sqrt{\aleph +4g^{2}N^{2}}$  & any  & ${\displaystyle\frac{1}{2}}$\\
& & & & & \\
\hline
& & & & & \\
$|2gNQ|$ & any &  any  &  ${\displaystyle\pm\frac{\aleph+4g^{2}N^{2}}{2g}}$  & any  & ${\displaystyle\frac{1}{4}}$\\
& & & & & \\
\hline\hline
\end{tabular}
\end{center}

\caption[]{\footnotesize In this table we represent 
the different combinations of 
values for the parameters $M,N,Q,P,\aleph$ of the general RN-TN-$aDS$ 
solution Eq.~(\ref{eq:RNTNaDS}) for which there are Killing spinors and 
the fraction of supersymmetry preserved.
The first two cases correspond locally to $aDS$. The last two cases are
the two general solutions of the constraint and Bogomol'nyi bound
equations and admit different limits with the same amount of supersymmetries
preserved. In particular, the third case preserves the same amount
of supersymmetry in the particular cases $Q=0$, $N=0,\aleph=+1$ (for any $Q$)
and $N=\pm 1/2g,\aleph=-1$. The fourth case preserves the same amount
of supersymmetry in the cases $Q=0$,  $N=0$ (for any $Q$) and $g=0$. 
In this last case, electric-magnetic invariance is preserved and $Q$ can
be substituted by $\sqrt{Q^{2}+P^{2}}$.}
\label{tab-RNTNADS}
\end{table}




\subsection{Supersymmetry of KN-TN-$aDS$ Solutions}

We choose the Vierbein 1-forms

\begin{equation}
\begin{array}{rcl}
e^{0} & = & {\displaystyle\frac{\lambda^{1/2}}{R(r,u)}
          \left[
             dt -\left(2Nu-a(\aleph^{2} -u^2)\right)d\varphi
          \right]} \, , \\
& & \\
e^{1} & = & 
{\displaystyle\frac{R(r,u)}{\lambda^{1/2}}dr} \, ,\\
& & \\
e^{2} & = & {\displaystyle\frac{ R(r,u)}{{\cal S}^{1/2}(u)}du} \, , \\
& &  \\
e^{3} & = & 
{\displaystyle\frac{{\cal S}^{1/2}(u)}{ R(r,u)}
\left[ (r^2+N^2+\aleph^{2} a^2)d\varphi \,+\, a dt \right]} \, ,
\end{array}
\label{eq:KNTNaDS4bein}
\end{equation}

%
%

\noindent on which the field strength components read

\begin{equation}
\begin{array}{rcl}
F_{01} & = & R(r,u)^{-4}
           \left[
              Q(r^2 -(N+au)^2 ) \,-\, 2Pr (N+au)
           \right] \; , \\
& & \\
F_{23} & = & -R(r,u)^{-4}
           \left\{
              P(r^2 -(N+au)^2 ) \,+ \, 2Qr(N+au)
           \right\} \; , \\
\end{array}
\end{equation}
%
%

We only need to calculate

\begin{equation}
  \begin{array}{rcl}
C_{0101} &=& -2R^{-6}\left[
               M(r^{3}-3rX^{2})
              +N(\aleph -\aleph^{2}a^{2}g^{2}+4g^{2}N^{2})(3r^{2}X-X^{3})
              -Z^{2}(r^{2}-X^{2})
             \right] \; ,\\
 & & \\
C_{0123} &=& 2R^{-6}\left[
               M(3r^{2}X-X^{3})
              +N(\aleph -\aleph^{2}a^{2}g^{2}+4g^{2}N^{2})(3rX^{2}-r^{3})
              -2Z^{2}rX
             \right] \; , \\
& &  \\
\nabla_{1}F_{01} &=& -2R^{-7}\lambda^{1/2}\left[
                       Q(r^{3}-3rX^{2})-P(3r^{2}X-X^{3})
                     \right] \; ,\\ 
& & \\
\nabla_{2}F_{01} &=& -2aR^{-7}{\cal S}^{1/2}\left\{
                      P(r^{3}-3rX^{2})+Q(3r^{2}X-X^{3})
                     \right\} \; ,\\
  \end{array}
\end{equation}

\noindent where we used the abbreviation $X=N+au$. As 
in the RN-TN-$aDS$ case the other components of the integrability
condition turn out to be proportional to the $01$ component.
{}From this one obtains the constraint and generalization of the Bogomol'nyi bound

\begin{equation}
\begin{array}{rcl}
0 & = & g\left[ MP+NQ(\aleph -\aleph^{2}a^{2}g^{2}+4g^{2}N^{2}) \right]\, ,\\
& & \\
0 & = & {\cal B}_{+}{\cal B}_{-}\, ,\\
\end{array}
\end{equation}

\noindent where now

\begin{equation}
\begin{array}{rcl}
{\cal B}_{\pm} & \equiv &
M^{2} +N^{2}(\aleph -\aleph^{2}a^{2}g^{2}+4g^{2}N^{2})
-\left[
(\aleph +\aleph^{2}a^{2}g^{2}+6g^{2}N^{2})
\right.
\\
& & \\
& & 
\left.
\pm
2g\sqrt{a^{2}(1+\aleph -\aleph^{2}) 
-N^{2}\left(\aleph -3\aleph^{2}a^{2}g^{2} +3g^{2}N^{2}\right)}
\right]Z^{2}\, ,\\
\end{array}
\end{equation}

The fact that the bound factorizes into the product ${\cal B}_{+}{\cal
  B}_{-}$ is difficult to see directly from the calculation but easy
to deduce from the results we will find in the general PD case.  It
can be checked that the (saturated) bound obtained is exactly the
same, when $N=0$, as the one given by Caldarelli and Klemm in
Ref.~\cite{art:CK}.

We can now try to analyze different solutions to these two equations.
This is a very complex problem and it would only make sense to explain
in detail a classification of the solutions if the different classes
had different amounts of unbroken supersymmetry. However, in all the
cases that we have been able to analyze we have not found any single
supersymmetric solution with $a\neq 0$ preserving $1/2$ of the
supersymmetries. In fact, adding angular momentum to the RN-TN-$aDS$
solutions that do preserve $1/2$ of the supersymmetries always seems
to break an another half leaving only $1/4$ unbroken.

For instance, the solution with $M=Q=0, P=\pm (2g)^{-1}(\aleph
-\aleph^{2}a^{2}g^{2}+4g^{2}N^{2}),\aleph=\pm 1$ preserves $1/2$ with
$a=0$ and only $1/4$ with $a\neq 0$. The same effect takes place
in all the instances studied.

\subsection{Supersymmetry of Topological RB Solutions}
\label{sec-susy-RB}

To check supersymmetry of the topological RB solutions we only need

\begin{equation}
C_{0101} = 
\textstyle{\frac{1}{3}}\left\{K^{2}-\aleph L^{2}\right\}= 2g^{2} \; ,
\end{equation}

\noindent since the vector field strengths is covariantly constant.

The integrability condition then reads

\begin{equation}
g\left[
     g\mathbb{I} 
  \,-\,\alpha\gamma^{01}i\sigma^{2}
  \,+\,\beta \gamma^{23}i\sigma^{2}
\right] \epsilon \;=\; 0 \; .
\end{equation}

Obviously, for $g=0$ one finds Robinson-Bertotti which will not break
any supersymmetry. When $g\neq 0$ however, one finds, just by taking 
the determinant of the above equation, that one has to satisfy

\begin{equation}
(g\pm\beta )^{2}\,+\,\alpha^{2} \,=\, 0 
   \hspace{.5cm}\rightarrow\hspace{.5cm}
\left\{
\begin{array}{lcl}
\alpha &=& 0  \\
\beta  &=& \pm g
\end{array}
\right.
\end{equation}

\noindent which then break half of the available supersymmetry. 
Plugging the above equations into Eq.~(\ref{eq:RobBertEis}), one finds
that

\begin{equation}
\aleph \,=\, -1 \hspace{.5cm},\hspace{.5cm}
K^{2} \;=\; 2 L^{2}\, ,
\end{equation}

\noindent which  means that $K^{2}=4g^{2}$ and $L^{2}=2g^{2}$. This is 
 the solution found in Ref.~\cite{kn:CKZ}. 

We could have found this solution also as the near-horizon limit of 
the $\aleph$ generalization of Romans' global monopole \cite{art:Ro}. In that 
case we have $P=\textstyle{\frac{\pm\aleph}{2g}}$ and with $\aleph =-1$
and all other charges vanishing  we find that there is a horizon 
at $2g^{2}r^{2}=1$. At this radius the solution can be approximated by

\begin{equation}
\begin{array}{rcl}
ds^{2} &=& 4g^{2}r^{2}dt^{2}-\frac{1}{4g^{2}r^{2}}dr^{2}
           -\frac{1}{2g^{2}}\left(
               d\theta^{2}\,+\, \sinh^{2}(\theta )d\phi^{2}
           \right) \; , \\
& & \\
F_{23} &=& -\frac{\pm 1}{2g}\cdot \left(\frac{1}{2g^{2}} \right)^{-1}
       \;=\; \mp g \; .\\
\end{array}
\end{equation}

\noindent which is just the supersymmetric RB-like solution discussed above.
We then see that we have supersymmetry enhancement at the horizon from
$1/4$ to $1/2$. Observe that he presence of  electric charge would have
meant the complete annihilation of supersymmetry at the horizon.

\subsection{Supersymmetry of the PD General Solution}
\label{sec-susy-PD}

As in the foregoing cases, one finds that all the components of the
integrability condition are equivalent, so we will only write down
the components of the Weyl tensor and the covariant derivative of the
vector field strength to calculate the integrability condition
in the $01$ direction.

\begin{equation}
\begin{array}{rcl}
C_{1010} & = & {\displaystyle\frac{-2(p+q)^{3}}{(1+p^{2}q^{2})^{3}}}
\left[
-\mathsf{M}(1-3p^{2}q^{2})
+\mathsf{N}(3pq -p^{3}q^{3})
+\mathsf{Z}^{2}(p-q)(1-p^{2}q^{2})
\right]\, ,\\
& & \\
C_{1023} & = & {\displaystyle\frac{-2(p+q)^{3}}{(1+p^{2}q^{2})^{3}}}
\left[
-\mathsf{M}(3pq-3p^{3}q^{3})
-\mathsf{N}(1-3p^{2}q^{2})
+\mathsf{Z}^{2}2pq(p-q)
\right]\, ,\\
& & \\
\nabla_{1}F_{01} & = & 
{\displaystyle\frac{2(p+q)^{2}{\cal Q}^{1/2}}{(1+p^{2}q^{2})^{7/2}} }
\left[
\mathsf{Q}(1-3p^{3}q +p^{5}q^{3}-3p^{2}q^{2})
+\mathsf{P}(3pq-p^{3}q^{3}+p^{2}-3p^{4}q^{2})
\right]\, ,\\
& & \\
\nabla_{2}F_{01} & = & 
{\displaystyle\frac{2(p+q)^{2}{\cal P}^{1/2}}{(1+p^{2}q^{2})^{7/2}} }
\left[
\mathsf{P}(1-3p^{3}q +p^{5}q^{3}-3p^{2}q^{2})
-\mathsf{Q}(3pq-p^{3}q^{3}+p^{2}-3p^{4}q^{2})
\right]\, ,\\
\end{array}
\end{equation}

Plugging these expressions into the integrability condition and
calculating the determinant, one finds that the following conditions
need to be satisfied in order for the solution to be supersymmetric

\begin{equation}
\begin{array}{rcl}
0 &=& g\left[ \mathsf{M}\mathsf{P}\,+\,\mathsf{N}\mathsf{Q}\right] \, ,\\
& & \\
0 &=& {\cal B}_{+}{\cal B}_{-}\, ,\\
\end{array}
\end{equation}

\noindent where, now

\begin{equation}
{\cal B}_{\pm} \equiv \mathsf{W}^{2}
-(\mathsf{E}\pm 2g\alpha^{1/2})\mathsf{Z}^{2}\, ,
\end{equation}

\noindent and we have defined $\mathsf{W}^{2}
=\mathsf{M}^{2}+\mathsf{N}^{2}$ and $\mathsf{Z}^{2}=
\mathsf{Q}^{2}+\mathsf{P}^{2}$.  One can check that these conditions
are invariant under the scalings in Eq.~(\ref{eq:PDscaling}) and they
give the integrability equations of the RN-TN-$aDS$ and KN-TN-$aDS$
cases after the redefinitions (\ref{eq:redefs}).

Again we find a constraint on the charges and a generalization of the
(saturated) Bogomol'nyi bound ${\cal B}_{\pm}=0$. The advantage of the
parameterization of the PD solution is, first of all, that the second
integrability condition factorizes completely and that ${\cal
  B}_{\pm}$ is extremely simple and is almost identical to the
standard bound for asymptotically flat, ungauged, $N=2,d=4$
supergravity solutions, being electric-magnetic duality-invariant and
invariant under gravito-electric-magnetic duality that rotates
$\mathsf{M}$ into $\mathsf{N}$ and vice-versa. These duality
invariances are broken by the constraint $g\left[
  \mathsf{M}\mathsf{P}\,+\,\mathsf{N}\mathsf{Q}\right] =0$ which is,
nevertheless invariant under {\it simultaneous} rotations with the
same angle

\begin{equation}
\left\{
\begin{array}{rcl}
\mathsf{M}^{\prime} & = & \cos{\theta}\mathsf{M} -\sin{\theta}\mathsf{N}\, ,\\
& & \\
\mathsf{N}^{\prime} & = & \sin{\theta}\mathsf{M} +\cos{\theta}\mathsf{N}\, ,\\
\end{array}
\right.
\hspace{2cm}
\left\{
\begin{array}{rcl}
\mathsf{Q}^{\prime} & = & \cos{\theta}\mathsf{Q} +\sin{\theta}\mathsf{P}\, ,\\
& & \\
\mathsf{P}^{\prime} & = & -\sin{\theta}\mathsf{Q}+\cos{\theta}\mathsf{P}\, .\\
\end{array}
\right.
\end{equation}

Actually, assuming that $g\neq 0$ one can eliminate completely the
constraint, getting a pair of equations

\begin{equation}
\left\{
\begin{array}{rcl}
\mathsf{M}^{2} & = & (\mathsf{E}\pm 2g\alpha^{1/2})\mathsf{Q}^{2}\, ,\\
& & \\
\mathsf{N}^{2} & = & (\mathsf{E}\pm 2g\alpha^{1/2})\mathsf{P}^{2}\, ,\\
\end{array}
\right.
\end{equation}

\noindent which hold even if some of these charges (but not $g$) vanish. 
These equations rotate into each other under the above duality
transformations.

The rotation parameter is always bounded above:

\begin{equation}
\alpha^{1/2}\leq  \pm\mathsf{E}/2g\, .  
\end{equation}

When this bound is saturated, then both $\mathsf{M}=0$ and
$\mathsf{N}=0$, while $\mathsf{Q}$ and $\mathsf{P}$ remain arbitrary.
This is always the case when $\mathsf{E}=0$.

Finally, the only supersymmetric solution with $\mathsf{Z}=0$ is
$aDS_{4}$.

A calculation of the rank of the integrability condition shows that 
all these configurations will generically break three-fourths of the 
available supersymmetries. This was to be expected from our results
in the KN-TN-$aDS$ case. On the other hand we have not been able to 
find any combination preserving up to $1/2$ of the available
supersymmetry which is not the RN-TN-$aDS$ solution.


\section{Conclusions}
\label{sec-conclusions}

In this letter we have presented new solutions which generalize the
already known topological black holes and the standard
Robinson-Bertotti solution. We have explored their supersymmetry
properties finding that generically they preserve only $1/4$ of the
supersymmetry. The only solutions that preserve $1/2$ are non-rotating
ones and the addition of angular momentum seems to break a further half
of the remaining supersymmetries.

A somewhat surprising result that deserves further study is the fact
that the most general family of supersymmetric solutions of this
theory (i.e.~the supersymmetric Plebanski-Demianski solutions) is
invariant under a continuous $SO(2)$ group of electric-magnetic
duality transformations. Had we not included in our study NUT charge
the existence of that symmetry would have passed completely unnoticed.
Its meaning is, however, obscure. After all, the charges that undergo
the duality rotation in its simplest, linear form, are not the
physical charges. In terms of the physical charges, the duality
transformations are very nonlinear.


\section*{Acknowledgments}

The work of N.A.-A., P.M.~and T.O.~have been supported in part by
the European Union TMR program FMRX-CT96-0012 {\sl Integrability,
  Non-perturbative Effects, and Symmetry in Quantum Field Theory} and
by the Spanish grant AEN96-1655. P.M. would like to thank Iberdrola
and the Universidad Aut\'onoma de Madrid for their support.

\appendix

\section{Obtaining the Topological KN-TN-aDS Metric from PD's}
\label{sec-KNTNADSfromPD}

Performing in Eq.~(2.11) the coordinate
change (see the analogous discussion in \cite{art:KMV})

\begin{equation}
\begin{array}{rclrcl}
q       & = & r\, ,                                                   & 
\tau    & = & t\,+\, \left[ a^{-1}N^{2}+a\aleph^{2}\right]\varphi \; ,\\
& & & & & \\
p       & = & N+au\, ,\hspace{2cm}                                    &
\sigma  & = & a^{-1}\phi \; ,                                         \\
\end{array}
\end{equation}

\noindent and the following redefinitions of the parameters
$\mathsf{M}=M$, $\mathsf{Q}=Q$, $\mathsf{P}=P$, 

\begin{equation}
\label{eq:redefs}
\begin{array}{rcl}
\mathsf{E} & = & 
\aleph\,+\, \aleph^{2}a^{2}g^{2}\,+\, 6g^{2}N^{2}\, , \\
& & \\
\mathsf{N} & = & 
N\left( \aleph\,-\, \aleph^{2}a^{2}g^{2}\,+\, 4g^{2}N^{2}\right)\; , \\ 
& & \\
\alpha     & = & 
a^{2}\left( 1+\aleph -\aleph^{2} \right)
\,-\, N^{2}\left(\aleph -3\aleph^{2}a^{2}g^{2} +3g^{2}N^{2} \right) 
+P^{2}\; ,\\
\end{array}
\label{eq:D&P2redefinitions}
\end{equation}

\noindent  we go from the PD metric to the KN-TN-aDS metric as written
down in Eqs.~(\ref{eq:KNTNaDS}).

Note that the choice of the redefinitions is largely dictated by the
factorizability of $\mathcal{P}$.


\end{document}